\documentclass[12pt,twoside]{article}
\usepackage{fleqn,espcrc1}

\usepackage{graphicx}
\usepackage[figuresright]{rotating}

\newcommand{\AmS}{{\protect\the\textfont2
  A\kern-.1667em\lower.5ex\hbox{M}\kern-.125emS}}

\title{Explosive Collisions at RHIC~?}

\author{Adrian Dumitru\address{Physics Department, Columbia Univ.,
        538 W. 120th Street, New York, NY 10027, USA\\}
        and
        Robert D. Pisarski\address{Physics Department, 
	Brookhaven National Laboratory,
        Upton, NY 11973 USA}}
       
\begin{document}

\maketitle

\begin{abstract}
Motivated by experimental results from RHIC, we suggest how
a condensate for the Polyakov loop might produce explosive behavior at the
QCD phase transition.  This is due to a rapid rollover 
of the condensate field below the transition temperature.
\end{abstract}

\vspace{.25in}
Many new results have poured forth from RHIC ($\sqrt{s} = 130$A~GeV).  
Several appear to be qualitatively
different from those at the SPS ($\sqrt{s} = 17$A~GeV).
``Dynamical'' fluctuations in the average
transverse momentum, $p_t$, are almost three times 
larger than at the SPS \cite{fluc}.  Observed HBT radii
are small, $\sim 5-7$~fm \cite{hbt}, and average transverse
boost velocities are large, $\sim .6$~c \cite{flow}.  In
particular, that the HBT radii $R_{out}/R_{side} \leq 1$,
as a function of pion pair $p_t: 200 \rightarrow 400 $~MeV, 
suggests ``explosive'' behavior \cite{hbt}.
Lastly, suppression of particles at high $p_t: 2 \rightarrow 6$~GeV
appears to be a clear diagnostic probe \cite{highpt,energyloss}.

Based upon
a condensate model \cite{rdp}, previously we predicted dynamical fluctuations
in the average transverse momentum \cite{adrp}.
Using known features of the condensate model \cite{adrp}, here we
suggest how it might produce explosive behavior at the QCD phase
transition, and affect the suppression of particles at high $p_t$.

At infinitely high temperature, by asymptotic freedom QCD is an ideal
gas of massless quarks and gluons.  Lattice data 
\cite{karsch} and effective theories \cite{qp}
suggest that from infinite temperature, on down to 
perhaps twice the transition temperature, $T_c$, it really is a
plasma: a nearly ideal gas of quarks and gluons, augmented
with ``thermal'' masses.

For temperatures of $2 T_c$ down to $T_c$, instead of quasiparticles,
it is more useful to think of a condensate \cite{rdp}.
In a nonabelian gauge theory without quarks,
the order parameter for the deconfining phase transition is the Polyakov
loop, $\ell$.  This is the trace of the
$SU(3)$ color Aharanov-Bohm phase factor
in the imaginary time direction.  By asymptotic freedom, the expectation
value of $\ell$, $\langle \ell \rangle = \ell_0$, 
is one at infinite temperature; from 't Hooft, it
is zero in the confined phase, $T < T_c$.  
Thus in going from $2 T_c$ down to $T_c$,
the condensate for the Polyakov loop evaporates.

In the condensate model, the pressure is determined by
a {\it potential} for $\ell$ \cite{rdp}, with 
\begin{equation}
p \sim \ell_0^4 \; p_{\rm ideal} \;\;\; , \;\;\; p_{\rm ideal} \sim T^4 \; ,
\end{equation}
where $p_{\rm ideal}$ is the pressure for an ideal gas of quarks and gluons.
In the pure gauge theory, where the quarks are quenched, 
the lattice finds that at $2 T_c$, 
$p/p_{\rm ideal} \sim .8$ \cite{karsch}, so maybe $\ell_0$ $\sim .95$.

Lattice data for $p/p_{\rm ideal}$  
can then be used to fit $\ell_0(T)$.
If $e$ is the energy density,
\begin{equation}
\frac{e - 3 p}{T^4} \sim T \; \frac{\partial \ell_0^4}{\partial T}
= 4 T \ell_0^3 \; \frac{\partial \ell_0}{\partial T} \; .
\end{equation}
In the pure gauge theory, 
the lattice finds that there is a sharp ``bump'' in $(e - 3p)/T^4$
above $T_c$.  This bump indicates that the transition occurs in a relatively
narrow region of temperature.  
In the condensate model, this reflects a rapid change in the
potential for $\ell_0$.

Although the lattice data is less reliable with dynamical
quarks, it suggests that the pressure 
is ``flavor independent'' \cite{karsch}.  In a mean field analysis
of the condensate model, this implies that the $\ell$-potential
is determined mainly by the gluons --- for which there is good lattice data.
Without quarks, the deconfining transition is weakly first order.
We make the most conservative assumption, that quarks wash out the
deconfining transition, to leave only crossover.  Even so, with
dynamical quarks, a bump in $(e - 3p)/T^4$ is still evident \cite{karsch},
indicating a rapid change in the $\ell$-potential near $T_c$.

For heavy ion collisions at high energies, 
the central region exhibits rapid longitudinal expansion along
the beam direction.  For an ideal gas,
with boost invariant Bjorken expansion the temperature falls
as $T \sim 1/\tau^{1/3}$, where $\tau$ is the proper time.  

If the transition
were strongly first order (as occurs for four or more colors), 
a mixed phase exists, and a hydrodynamic analysis indicates that 
the system lasts for a long time at $T_c$ \cite{ed}.  Because of this,
the HBT ratio $R_{out}/R_{side}$ is significantly larger than one, $\sim 2$ 
\cite{dirk}.  Thinking of the firetube in the central region as a log,
this is the ``burning log'' scenario \cite{dirk}.  A strong first
order transition implies that there is a mixed phase of bubbles at $T_c$.
For slow expansion, bubbles with 
$\ell_0 = 0$ coexist, at equal pressure, with those with $\ell_0 \neq 0$
\cite{owe2}.

If the transition is crossover, 
and the potential for $\ell$ changes
slowly, then the system evolves smoothly through $T_c$, with a 
gradual shift of $\ell_0$ from $\sim 1$ to $\sim 0$.  
This is a ``smoldering log'', with $R_{out}/R_{side} > 1$.

If the transition is crossover (or weakly first order), {\it and} the
potential for $\ell$ changes {\it rapidly}, then the system 
quickly passes through $T_c$.  Suddenly, the system finds itself
below $T_c$, at a nonzero value of $\ell_0$, which
is no longer the minimum.  It then rolls down, oscillating about zero,
until it settles to $\ell_0 = 0$.  How this roll down occurs is
described by the following equation:
\begin{equation}
Z_0 \frac{\partial^2}{\partial \tau^2} \, \ell
+ \gamma_\ell \frac{\partial}{\partial \tau} \, \ell
- Z_s \nabla^2 \ell + {\cal V}'(\ell) + h_\Phi \Phi^2  \ell = 0.
\end{equation}
Here $\nabla^2$ is the Laplacian in the spatially
transverse and rapidity directions, 
and ${\cal V}(\ell)$ is the potential for $\ell$.  
The Polyakov loop is coupled to a field $\Phi$ for Goldstone
bosons ($\pi$, $K$, $\eta$, $\eta'$) through 
a term $\sim h_\Phi \Phi^2 \ell$ \cite{adrp}.  In the
condensate model, the potential ${\cal V}(\ell)$ is completely
fixed by the pressure.  
The coefficient for spatial variations in the Polyakov
loop, $Z_s$, was taken from that for $SU(3)$ Wilson lines,
$Z_s \sim 1/g^2$ \cite{adrp}; actually, $Z_s \sim g^4$.  We also assumed
a Lorentz invariant form, $Z_0 = Z_s$, although probably
$Z_0 \neq Z_s$.

In \cite{adrp} we analyzed this equation in the Hartree approximation,
where $\Phi^2 \rightarrow \langle \Phi^2 \rangle$, which neglects
collisions amongst the produced pions.  The condensate field $\ell$
was assumed to be spatially homogeneous, so that only its variation
in time mattered.  
We took $\gamma_\ell = 0$; this neglects the effects of longitudinal
expansion, which contributes $1/\tau$ to $\gamma_\ell$, and dissipation.
Above $T_c$, the potential is sharp, and $\ell$ is trapped in a
minimum with $\ell_0 \neq 0$.  For a window of temperature which is
extremely narrow, $\Delta T \approx 2 \% T_c$, the potential changes
suddenly into that for the symmetric phase: $\ell$ rolls down,
and oscillates about zero.  The initial energy density, which is
stored entirely in the $\ell$ field, is then pumped into the production
of pions.
As the initial energy density is large, the produced pions have
large average transverse momentum, $\sim 5 f_\pi = 500$~MeV.  We
note that while this agrees with the experimentally measured 
result \cite{flow}, in the present model this value cannot be tuned.  
The system is underdamped, and the average
field oscillates several times about zero.
These oscillations produce
relatively large fluctuations about the average transverse momentum, $p_t$ 
\cite{adrp}, on the order of $\sim 10\%$.  
This scenario is like reheating after inflation, or a quench in 
Disoriented Chiral Condensates (DCC's) \cite{adrp,dcc}.  
It appears unlikely, however, that many coherent oscillations of
an $\ell$-field, sloshing back and forth, generates explosive behavior. 

The STAR results suggest an alternate scenario.  Perhaps, due
either to rapid longitudinal expansion, or to an intrinsic
term $\gamma_\ell \neq 0$, one is in a regime where the effects of
damping cannot be neglected.  
In this case, the potential evolves quickly, but because of
the damping, $\ell_0$ oscillates 
only a few times.  It is conceivable --- but
not guaranteed --- that in this case particle production is
explosive.  Particles are emitted from a shell \cite{greiner},
with $R_{out}/R_{side} \leq 1$.  

A semiclassical calculation of particle production in the condensate
model appears reasonable.  For a large number of colors, $N_c$,
the potential ${\cal V}(\ell) \sim N_c^2$ \cite{rdp}, while
$Z_0$, $Z_s$, $\gamma_\ell \sim 1$; thus a semiclassical calculation
is nominally valid to $\sim 1/N_c^2 \sim 10\%$ for $N_c = 3$.  
Getting spectra which are exponential in transverse mass is probably
easy \cite{dirkqm}; getting relative ratios right, 
such as $K/\pi \sim 0.2$, is far from obvious.  
Explosive
behavior should also produce a characteristic electromagnetic spectrum,
such as in direct photons \cite{boy}.

Both scenarios produce relatively large
fluctuations about the average transverse momentum, which
are measurable on an Event-by-Event basis \cite{event}.  In the underdamped
scenario, production is from causally connected regions, of
size $\sim 1$~fm \cite{adrp}.  
Due to Poisson statistics,
the fluctuations from each domain are then reduced by
the inverse square root of the number of domains in one unit of rapidity;
If in one unit of
rapidity there are $\sim 300$ domains, the fluctuations are
$\sim 10\%/\sqrt{300} \sim .6 \%$.  Further, 
smaller rapidity bins should increase the dynamical fluctuations;
for a bin $\Delta y=0.25$,
fluctuations increase by a factor of two, to $\sim 1.2 \%$.
This increase in fluctuations on smaller scales is seen in the
cosmic microwave background.  In the damped scenario, 
if the time scale for
damping is much less than the spatial correlation length, then fluctuations
may be essentially constant with the size of the bin.  

CERES at the SPS finds dynamical fluctuations of the 
mean $p_t$ $\leq 3\%$; STAR finds fluctuations $\sim 8\%$ \cite{fluc}.  
Regardless of which scenario applies, the condensate model 
predicts that the fluctuations are concentrated about the mean, and
not due to the tails of the distribution in $p_t$.  
Thus the experimental signal should not be significantly
affected if one cuts on a relatively narrow bin in momentum.  
STAR included all particles with $.1 < p_t < 1.5$~GeV;  in
this model the fluctuations are the same for particles
with $.3 < p_t < .7$~GeV.  Binning in a narrower region in $p_t$
also serves to distinguish it from 
other models.  DCC's \cite{dcc}
or a chiral critical point \cite{crit} produce 
fluctuations at low $p_t$; jets give fluctuations at high $p_t$.

We can also make qualitative predictions about energy loss \cite{energyloss};
in perturbation theory, the energy loss for a fast particle with
transverse momentum $p_t$ is
\begin{equation}
dE/dx \sim \alpha_s m_{\rm Debye}^2/\lambda_g \sim \ell_0^2 \; ,
\end{equation}
where $\alpha_s(p_t)$ is the strong coupling constant, $m_{\rm Debye}$
is the Debye mass, and $\lambda_g$ is the gluon mean free path.  
In the condensate model, $m_{\rm Debye}^2 \sim \ell_0^2$, and so
vanishes at $T_c$.  This is also seen directly from the lattice 
data \cite{karsch}.  Assuming that $\lambda_g$ is finite at
$T_c$ gives the dependence on $\ell_0$ given above.  
In QCD, energy loss becomes small as $\ell_0$ does.  

PHENIX and STAR presented results which appeared to show suppression of
particles at high $p_t: 2 \rightarrow 6$~GeV \cite{highpt}.  This is 
seen even by comparing peripheral to central collisions, as
then no model dependent assumptions need to be made.  This suggests
that at $\sqrt{s} = 130$A~GeV, one is in a regime in which
$\ell_0 \neq 0$. 
At the SPS, no such suppression is observed,
which suggests that $\ell_0 \approx 0$ at $\sqrt{s} = 17$A~GeV.
(At RHIC, perhaps the largest $\ell_0$ is still
$< 1$, with $\ell_0 \approx 1$ only at LHC?)  

The outstanding question is then how these various signals change
between $\sqrt{s} = 17$A~GeV and $130$A~GeV; is it a sudden 
transition, or gradual \cite{flow}?  If the explosive scenario
is correct, one might speculate that the transition region is relatively
narrow.  However, the relationship between temperature and $\sqrt{s}$
may be far from linear.  In saturation models \cite{sat}, for example,
$T \sim (\sqrt{s})^{0.2}$, so a relatively narrow region in $T$ 
corresponds to a broad region in $\sqrt{s}$.

\end{document}